\documentclass[aps,prl,preprint, titlepage,showpacs]{revtex4}

\usepackage{graphicx}% Include figure files

\usepackage{dcolumn}% Align table columns on decimal point

\usepackage{bm}% bold math

\begin{document}

\title{Test of the Stokes-Einstein relation in a two-dimensional
Yukawa liquid}

\author{Bin Liu}
\author{J. Goree}
\affiliation{Department of Physics and Astronomy, The University
of Iowa, Iowa City, Iowa 52242}

\author{O. S. Vaulina}
\affiliation{Institute of High Temperatures, Russian Academy of
Sciences, 13/19 Izhorskaya St, Moscow 127412, Russia}

\date{\today}

\begin{abstract}

The Stokes-Einstein relation, relating the diffusion and viscosity
coefficients $D$ and $\eta$, is tested in two dimensions. An
equilibrium molecular-dynamics simulation was used with a Yukawa
pair potential. Regimes are identified where motion is diffusive
and $D$ is meaningful. The Stokes-Einstein relation, $D \eta
\propto k_{B}T$, was found to be violated near the disordering
transition; under these conditions collective particle motion
exhibits dynamical heterogeneity. At slightly higher temperatures,
however, the Stokes-Einstein relation is valid. These results may
be testable in strongly-coupled dusty plasma experiments.

\end{abstract}

\narrowtext

\maketitle

Two-dimensional systems in crystalline or liquid
states~\cite{Strandburg88} are of interest in various fields of
physics. Monolayer particle suspensions can be formed in colloidal
suspensions~\cite{Murray90} and dusty plasmas~\cite{Nunomura00}.
Electrons on the surface of liquid helium form a 2D Wigner
crystal~\cite{Grimes79}. Ions in a Penning trap can be confined
as a single layer of a one-component plasma
(OCP)~\cite{Mitchell99}. Magnetic flux lines in 2D
high-temperature superconductors form patterns of hexagonally
correlated vortices~\cite{Gammel87}. Granular flows of discrete
macroscopic particles can be confined to a 2D
layer~\cite{Jaeger1996}. Here we will be concerned with 2D
liquids, with no motion in the vertical direction. The
experimental system listed above actually have finite out-of-plane
motion, our result will be most applicable to systems such as
dusty plasma where the out-of-plane displacement is small compared
to the particle spacing. As in 3D liquids~\cite{Glotzer00}, 2D
liquids near the disordering transition are especially interesting
because they exhibit dynamical heterogeneities, where the
collective motion of the more mobile particles occurs in
strings~\cite{Reichhardt03}.

Our 2D system is composed of molecules or particles that interact
with a Yukawa pair potential. This is a soft potential that is
applicable in several fields including
colloids~\cite{Murray90,Reichhardt03}, dusty
plasmas~\cite{Nunomura00,Vaulina04}, and some
polyelectrolytes~\cite{Denton03} in biological and chemical
systems.

The Stokes-Einstein relation for 3D liquids
\begin{equation}\label{S-Eeq}
  D = k_{B}T/c\pi\eta R
\end{equation}
is an important hydrodynamic law relating the diffusion
coefficient $D$ of a Brownian particle and the fluid shear
viscosity $\eta$. It states that the diffusion of a Brownian
particle in a liquid is proportional to the liquid temperature
$k_{B}T$ and is inversely proportional to the viscosity of the
liquid. This combines the Einstein relation $D=k_{B}T/\xi$ for $D$
and the Stokes law $\xi=c\eta R$ for the frictional force $\xi$ on
a sphere in a fluid. Here, $c$ is a constant that depends on the
hydrodynamic boundary condition at the particle surface, and $R$
is the effective radius of the Brownian particle.

For 3D liquids, the validity of the Stokes-Einstein relation is
well known, even at the molecular level. Over a wide range of
temperature~\cite{Parkhurst1975}, the diffusion coefficient $D$
and the factor $k_{B}T/\eta$ have a fixed ratio as $k_{B}T$
varies. Only for deeply supercooled liquids does the
Stokes-Einstein relation appear to fail~\cite{Ernst1990}; such
liquids typically exhibit a spatial
heterogeneity~\cite{Glotzer00}. The Stokes-Einstein relation is
so trusted in 3D that it is routinely used to compute $D$ from a
measurement of $\eta$. However, the situation in 2D is very
different.

For 2D liquids, unlike 3D liquids, tests of the validity of the
Stokes-Einstein relation are lacking. Indeed, some earlier results
suggest that the Stokes-Einstein relation should not be valid at
all in 2D because the transport coefficients $D$ and $\eta$ are
themselves not valid. These earlier results came from a 2D
molecular dynamics (MD) simulation~\cite{Alder70} and kinetic and
mode-coupling theories \cite{Ernst70,Dorfman70}. They predict
that, unlike the case of 3D, the long-time tails of the velocity
(VACF) and stress autocorrelation functions (SACF) have a slow
$1/t$ decay. A decay this slow precludes the use of Green-Kubo
relations to calculate $D$ and $\eta$, and suggests that these
transport coefficients are meaningless, at least in the context of
equilibrium liquids. However, there is no universal agreement that
$D$ and $\eta$ are meaningless in 2D. The shear viscosity
coefficient in 2D has now been shown to be meaningful, at least
for soft potentials, in a significant number of
simulations~\cite{Bin05, Ferrario97, Gravina95, Hoover95}, and
an experimental measurement~\cite{Vova04} of a 2D viscosity has
been reported. The diffusion coefficient in 2D, on the other hand,
has not been demonstrated to be meaningful over a wide range of
density or temperature. As a test of this, motion is deemed to be
diffusive, and diffusion coefficients are meaningful, only if the
VACF decays faster than $1/t$, or if the Einstein relation ${\rm
MSD}\propto t$ is obeyed. Here, ${\rm MSD}$ is the time series of
mean-square displacement of particle positions.
Simulations~\cite{Zangi04, Donna98, Huley95}, using a $r^{-12}$
pair potential and the Einstein relation, for example,
successfully yielded a diffusion coefficient, but only near
disordering transition. Identifying the conditions where a 2D
diffusion coefficient is meaningful is our first goal. We do this
two ways, using the Green-Kubo method and the Einstein relation.
The expressions we used for the VACF and Green-Kubo relations were
Eqs.~(7.2.1) and (7.2.8), respectively, in Ref.~\cite{Hansen86},
where we replace the coefficient 3 with 2, to reflect the number
of dimensions.

Our second goal is to test the Stokes-Einstein relation for 2D
liquids. For the regimes where $D$ is valid we will compute the
product $D\eta$ and plot it versus temperature as an empirical
test of the 2D Stokes-Einstein relation,
\begin{equation}\label{S-Eeq-2D}
D=k_{B}T/c_{2D}\pi\eta.
\end{equation}
Equation (\ref{S-Eeq-2D}) is analogous to Eq.~(\ref{S-Eeq}),
except that it does not contain the particle radius $R$ in the
denominator due to the different dimensionality of $\eta$ in 2D.
Finding a linear scaling of $D\eta$ on $k_{B}T$ would be an
empirical verification of the validity of the Stokes-Einstein
relation, while a deviation from linear scaling would indicate a
violation. In performing this test, it is important that we
obtained values of $D$ and $\eta$ independently.

We use a 2D Yukawa system. Particles are present only in a single
monolayer, and they interact with a Yukawa potential energy, which
for two particles of charge $Q$ separated by a distance $r$ is
$U(r)=(Q^{2}/4\pi\epsilon_{0}r)\exp(-r/\lambda_{D})$. This
potential changes gradually from a long-range Coulomb repulsion to
a hard-sphere-like repulsion as the screening parameter
$\kappa=a/\lambda_{D}$ is increased. Here, $\lambda_{D}$ is a
screening length, $a=(n\pi)^{-1/2}$ is the 2D Wigner-Seitz
radius~\cite{Kalman04}, and $n$ is the areal number density of
particles.  A Yukawa system is characterized by two dimensionless
parameters $\kappa$ and the Coulomb coupling parameter
$\Gamma=Q^{2}/4\pi\epsilon_{0}akT$. We will use $\Gamma^{-1}$ as a
dimensionless temperature. The phonon spectrum of a 2D Yukawa
system is characterized by a frequency
$\omega_{pd}=(Q^{2}/2\pi\epsilon_{0}ma^{3})^{1/2}$~\cite{Kalman04}.
The disordering transition for a 2D Yukawa system occurs at
$\Gamma=137$ for $\kappa=0$~\cite{Grimes79}. At our value of
$\kappa=0.56$, the transition is at
$\Gamma\approx145$~\cite{Hartmann05}, and we denote the
corresponding melting temperature as $T_{m}$.

We performed an MD simulation to calculate $\eta$ and $D$. The
equations of motion for 1024 or 4096 particles were integrated
using periodic boundary conditions. A Nos{\' e}-Hoover thermostat
was applied to achieve a constant $T$. As a test, we ran
simulations with the thermostat turned on or off, and verified
that the thermostat does not affect $\eta$ or $D$, for any values
of $\Gamma$. The screening parameter was $\kappa=0.56$. We
recorded particle positions and velocities, and then we calculated
velocity and stress correlation functions and the mean-square
displacement. The system was in equilibrium, and it had no
macroscopic velocity shear or density gradient. There was no
friction term in the equation of motion. Examples of particle
trajectories for $\Gamma=104$ and $129$ are shown in
Fig.~\ref{trajectory}. Data for MSD and correlation functions were
averaged over multiple runs with different initial conditions.
Further details of the simulation can be found in
Ref.~\cite{Bin05}.

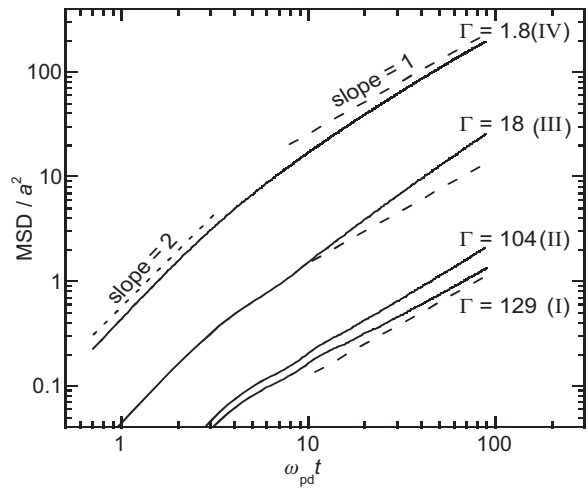
\begin{figure}[p]
\caption{\label{MSD} Mean-square displacement (MSD). Ballistic
motion at small $t$ has ${\rm MSD}\propto t^{2}$, with a slope of
2 on this log-log plot. Motion at later times is either diffusive,
${\rm MSD}\propto t$ as indicated by a slope of 1, or
non-diffusive as in regime III. By this test, we found that motion
is diffusive in regimes I, II and IV, but not III. A separate
test, using the VACF, is consistent with this result.}
\end{figure}

To meet our first goal, we identify regimes where the transport
coefficients $D$ and $\eta$ are meaningful.
Previously~\cite{Bin05} we found that $\eta$ was meaningful, as
indicated by a SACF that decayed faster than $1/t$, for a wide
range of $\Gamma$. Here, we report results for the diffusion
coefficient $D$, where the results are more complicated. We
applied two tests to determine whether motion is diffusive. First,
if ${\rm MSD}\propto t^{1}$ is satisfied at times that are long
(compared to the short-term ballistic motion, which has an
exponent of 2, ${\rm MSD}\propto t^{2}$), then the Einstein
relation is satisfied, indicating diffusive motion. Results for
this test for selected values of $\Gamma$ are shown as a time
series for ${\rm MSD}$ in Fig.~\ref{MSD}. The time series has a
duration that exceeds, by approximately a factor of 2, the ratio
of the box width and sound speed. We test the last portion of this
time series, for an interval of $\approx37\omega_{pd}^{-1}$, to
determine whether the exponent is equal to 1. Second, if the
long-time tail in the VACF decays faster than $1/t$, then
Green-Kubo theory indicates that the motion is diffusive. We found
that these two tests always agreed. Details of the VACF data will
be reported elsewhere.

We identified four different temperature regimes, and we
characterized them according to whether the motion is diffusive or
not. All four regimes have $\Gamma<145$, i.e., they are all in the
liquid state and have $T>T_{m}$. In regime I ($124<\Gamma<145$)
the motion is diffusive; this regime is near the disordering
transition. Motion is also diffusive in regime II
($88<\Gamma<124$), and the non-ideal gas regime IV ($\Gamma<5$).
However, motion was not diffusive in regime III ($5<\Gamma<88$).
That a regime exists where motion is non-diffusive is not
surprising, because earlier studies of 2D liquids predicted they
would be non-diffusive due to the long-time tail in the VACF,
which arises from hydrodynamic
modes~\cite{Alder70,Ernst70,Dorfman70}. What is surprising is
that we found other regimes, for our frictionless atomic system,
that are diffusive.

If a friction term were added to the equation of motion to model
drag due to a solvent the motion might more likely be diffusive.
L{\"{o}}wen~\cite{Lowen92} used a different simulation method,
Brownian dynamics, for a 2D Yukawa system with a large friction.
The equation of motion in that method lacks an inertial term, and
is therefore suitable for a strongly dissipated system. They
reported finding diffusive motion not only in our regimes I and
II, but also for one data point in our regime III. Their values
for $D$ were systematically smaller than ours by a factor of 2.3
in regime II.

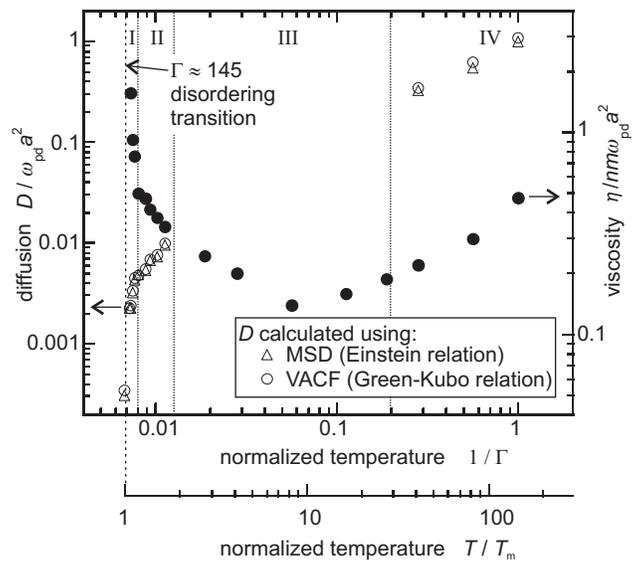
\begin{figure}[p]
\caption{\label{coefficients} Diffusion and viscosity
coefficients. Diffusion coefficients (open symbols) were computed
from the same simulation data using two methods, the Green-Kubo
relation and the Einstein relation. For comparison to the
diffusion coefficient, the viscosity coefficient data from
Ref.~\cite{Bin05} are shown as solid circles; these were computed
using a Green-Kubo relation. The disordering transition for our
$\kappa=0.56$ occurs at $\Gamma\approx145$, based on the
simulation results of Ref.~\cite{Hartmann05}.}
\end{figure}

Our two methods of computing $D$ yield values that agree, as shown
in Fig.~\ref{coefficients}. Note that $D$ increases with
temperature, and the rate of increase is very high in regime I,
near the disordering transition at $T_{m}$.

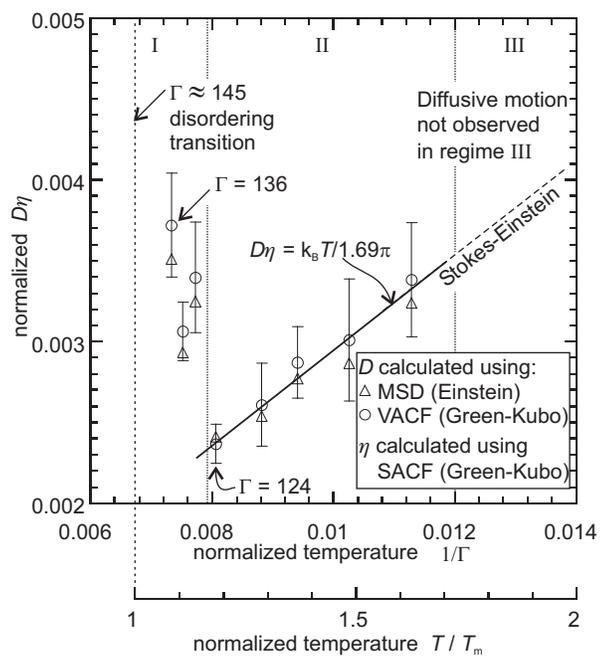
\begin{figure}[p]
\caption{\label{S-Etest}~Test of the Stokes-Einstein relation.
Linear scaling $D\eta\propto T$ in regime II indicates the
validity of the Stokes-Einstein relation. However, the deviation
from the linear scaling for $\Gamma>124$ indicates a violation of
the Stokes-Einstein relation in regime I, which is near the
disordering transition. Diffusive motion was not observed in
regime III, so that the relation is not testable there. The
vertical axis is normalized by $nm\omega_{pd}^2 a^{4}$.
Extrapolating the Stokes-Einstein line shown here into regime IV
(not shown), we found rough agreement with simulation data even
though the applicability of the Stokes-Einstein relation in
non-ideal gas is not expected.}
\end{figure}

To meet our second goal of testing the Stokes-Einstein relation,
we compute the product $D\eta$ and examine its scaling with
temperature. Results, shown in Fig.~\ref{S-Etest}, reveal that
$D\eta$ varies almost linearly with temperature in regime II, but
not regime I. We therefore conclude that the Stokes-Einstein
relation is obeyed in regime II but not in regime I.

We can thus summarize our two chief results. First, the diffusion
coefficient is meaningful in both regimes I and II but not regime
III. Second, the Stokes-Einstein relation is valid in regime II,
but it is violated in regime I.

We found the value of the coefficient in Eq.~(\ref{S-Eeq-2D})
relating $D$, $\eta$, and $k_{B}T$. Fitting the data in
Fig.~\ref{S-Etest} to a straight line passing through the origin
yields $c_{2D}=1.69\pm0.10$. This coefficient has never been
reported, to our knowledge, for 2D, although its value is
well-known for 3D, where $c=4$ for stick or $c=6$ for slip
boundary conditions.

Verification of the Stokes-Einstein relation in regime II
indicates a coupling of diffusion and shear viscosity in 2D just
as it does in 3D simple liquids. For a macroscopic particle in a
fluid, particle diffusion is a result of Brownian motion, with a
friction that arises from the viscous dissipation of the
associated velocity field in the surrounding liquid. In a simple
liquid, validity of the Stokes-Einstein relation is an indication
of a coupling of individual particle motion to collective modes.
This coupling has been observed previously in 3D for
strongly-coupled Yukawa systems~\cite{Ohta00} and one-component
plasmas~\cite{Schmidt97}.

The violation of the Stokes-Einstein relation that we observed in
regime I, near the disordering transition, indicates a decoupling
of diffusive and viscous transport. We found the diffusion
coefficient is larger than would be expected if the
Stokes-Einstein relation were valid. We note that in 3D a
violation of the Stokes-Einstein relation is also observed in
supercooled liquids~\cite{Ernst1990,Dzugutov02}, and this
motivates us to suggest an explanation for our result in 2D.

The decoupling in regime I (near the disordering transition) is
likely related to structural and dynamical heterogeneities in the
liquids. As shown in Fig.~\ref{trajectory}, on a time scale of
tens of $\omega_{pd}$ the 2D liquid develops domains of localized
crystalline order with localized oscillatory motion that are
separated by stringlike structures of more mobile particles. For
3D supercooled liquids, it is believed that this type of
heterogeneity accounts for the failure of the Stokes-Einstein
relation~\cite{Dzugutov02}. Our results suggest that the same
effect is responsible in 2D.

The results we have reported for our test of the Stokes-Einstein
relation in a 2D liquid could be tested in a future experiment.
Monolayer suspensions of microsphere in a dusty plasma have a
Yukawa interaction as in our simulation, and a recently-developed
method has been reported for measuring the shear viscosity. There
is no obstacle to measure the self-diffusion coefficient as well.

We thank V. Nosenko and F. Skiff for helpful discussions. This
work was supported by NASA, DOE, and CRDF No. RU-P2-2593-MO-04.

\begin{figure}[p]
\caption{\label{trajectory} Particle trajectories from the
simulation. Regime I is near the disordering transition, and
regime II is at a slightly higher temperature. Regime I has a
dynamical heterogeneity; strings of mobile particles flow between
domains consisting of less-mobile caged-particles. In regime II
($\Gamma=104$) the domains are smaller and there is a less
distinctive difference between the caged trajectories and the
trajectories of particles bounding the domains. Trajectories are
shown for a time interval $\omega_{pd}\Delta t\approx37$. Only a
portion of the simulation box is shown here. The full simulation
box was $113.98a\times98.16a$ with 4096 particles for
$\Gamma=129$, and $56.99a\times49.08a$ with 1024 particles for
$\Gamma=104$.}
\end{figure}

\end{document}